\documentclass[12pt,preprint,showeys,preprintnumbers,amsmath,amssymb]{revtex4}
\usepackage{amsfonts}
\usepackage[latin1]{inputenc}
\usepackage[OT1]{fontenc}
\usepackage{graphicx}
\usepackage{dcolumn}
\usepackage{longtable}
\usepackage{rotating}

\begin{document}

\large
\noindent
{\bf Symmetry-adapted formulation of the combined G-particle-hole Hypervirial equation and Hermitian Operator method}

\vskip 5 mm

\normalsize
\noindent
{\bf Diego R. Alcoba$^{a,b,*}$, Gustavo E. Massaccesi$^{c}$, 
Ofelia B. O\~na$^{d}$, Juan J. Torres$^{e}$, Luis Lain$^{f}$, Alicia Torre$^{f}$}

\noindent
\scriptsize

\noindent
$^{a}${\it Departamento de F\'{\i}sica, Facultad de Ciencias Exactas 
y Naturales, Universidad de Buenos Aires. Ciudad Universitaria, 1428 
Buenos Aires, Argentina}\\ 
$^{b}${\it Instituto de F\'{\i}sica de Buenos Aires, Consejo Nacional 
de Investigaciones Cient\'{\i}ficas y T\'ecnicas. Ciudad Universitaria, 
1428 Buenos Aires, Argentina}\\
$^{c}${\it Departamento de Ciencias Exactas, Ciclo B\'asico Com\'un,
Universidad de Buenos Aires, Ciudad Universitaria, 1428 Buenos Aires, Argentina}\\
$^{d}${\it Instituto de Investigaciones Fisicoqu\'{\i}micas Te\'oricas 
y Aplicadas, Universidad Nacional de La Plata, CCT La Plata, Consejo 
Nacional de Investigaciones Cient\'{\i}ficas y T\'ecnicas.
Diag. 113 y 64 (S/N), Sucursal 4, CC 16, 1900 La Plata, Argentina}\\
$^{e}${\it Departamento de Ciencias Qu\'{\i}micas, Facultad de Ciencias Exactas, Universidad Andres Bello, Av.
República 275, Santiago de Chile, Chile}\\
$^{f}${\it Departamento de Qu\'{\i}mica F\'{\i}sica, Facultad de 
Ciencia y Tecnolog\'{\i}a, Universidad del Pa\'{\i}s Vasco. Apdo. 
644 E-48080 Bilbao, Spain}\\

\vskip 1mm

\normalsize
\noindent
\rule{155mm}{0.2mm}
\noindent
ABSTRACT\\
\noindent
\rule{155mm}{0.2mm}
\noindent
High accuracy energies of low-lying excited states, in molecular systems,
have been determined by means of a procedure which combines the G-particle-hole Hypervirial (GHV) equation 
method [Alcoba et al. Int. J. Quantum Chem. 109:3178 (2009)] and the Hermitian Operator (HO) 
one [Bouten et al. Nucl. Phys. A 202:127 (1973)]. This paper reports a suitable
strategy to introduce the point group symmetry within the framework of the combined GHV-HO method, what leads 
to an improvement of the computational efficiency. The resulting symmetry-adapted formulation has been applied to
illustrate the computer timings and the hardware requirements in selected
chemical systems of several geometries.
\\
\noindent
\rule{155mm}{0.2mm}


\noindent
\rule{20mm}{0.2mm}\\
\noindent
{\scriptsize
$^{*}$ Corresponding author. 
\vskip -4mm
\noindent
{\it E-mail address:} qfxaldad@lg.ehu.es}

\newpage

\normalsize
\noindent
{\bf 1. Introduction}
\vskip 2mm

All the fundamental electronic properties, including the energy, can be expressed as expectation
values of one- and two-electron operators. Therefore, they can be determined using only
the 2-order reduced density matrix (2-RDM) without recourse to the $N$-body wave function.
Both variational and non-variational approaches have been developed to the direct determination
of the 2-RDM for electronic systems. There is a large bibliography on this subject, which the interested reader
may find in the books of Davidson \cite{1} and Coleman and Yukalov \cite{2} as well as in
many proceedings and reviews \cite{3,4,5,6,7,8}. In the last years our interest has been focused on a non-variational 
method based on the iterative solution of the G-particle-hole hypervirial equation (GHV) \cite{9}, which results 
from the contraction of a particular case of the quantum Liouville equation \cite{10}. The accuracy of the results 
obtained with the GHV method when studying the ground state of molecular
systems at equilibrium geometry was excellent when compared with the equivalent
Full Configuration Interaction (FCI) quantities \cite{9,11,12,13}. However, the
study of the excited states is still a partially open question \cite{14,15}. 

Since the GHV method provides us with a good description of the ground states, we have recently studied the suitability 
 to combine this method with the Hermitian Operator (HO) method of Bouten {\it et al.} \cite{16,17} for computing excited 
state energies directly from the sole knowledge of the ground-state 2-RDM, or, equivalently, of the 
G-particle-hole matrix, which is obtained by solving the GHV equation  \cite{18}. Applications to molecular systems have 
shown that this combined GHV-HO method can yield accurate energy values not only for excited-states but also 
for some ground states in which the GHV method presents difficulties \cite{18,19,20}.

The aim of this work is to enhance the efficiency of the combined GHV-HO method by the exploitation
of molecular point group symmetry. Following recent work made within the framework of the GHV method \cite{21},
symmetry-related analysis of the matrices and matrix operations involved in the HO method is carried out.
This analysis leads to a symmetry-adapted formulation of the combined GHV-HO algorithm for Abelian groups which
generates significant computational savings in both floating-point operations and memory storage.

The paper is organized as follows. In the next section the notation, definitions and
general theoretical background of the GHV and HO methods are given.
In section 3 we describe the symmetry-adapted formulation of the GHV-HO method.
A number of statistics pertaining to the computational
cost of GHV-HO calculations are presented and analyzed for a set of molecules in section 4. Finally, the 
conclusions of this work are given in the
last section.

\vskip 7mm
\noindent
{\bf 2. Basic theoretical background}
\vskip 2mm
\noindent
{\bf 2.1. Notation and basic definitions}
\vskip 2mm

In what follows we will consider
pairwise-interacting systems composed of fixed number $N$
of electrons, whose Hamiltonian $\hat H$ may be written within second quantization formalism \cite{22} as  

\begin{equation}
\hat H\;=\;\frac 12\sum\limits_{pq;rs}\;^0{\rm H}^{pq}_{rs}\;a^{p \dagger}\,a^{q \dagger}\,a_{s}\,a_{r}
\end{equation}
where $a^{p \dagger}$ and $a_{r}$ are second quantization creation and annihilation operators, the indices
refer to members of a finite basis set of $2K$ orthonormal spin-orbitals, and $^0{\rm H}$ is a 2-order matrix
which collects the 1- and 2-electron integrals, $\epsilon^{q}_{s}$ and $\langle pq|rs\rangle$ respectively,
\begin{equation}
^{0}{\rm H}^{pq}_{rs}\;=\;\dfrac{\delta^{p}_{r}\;\epsilon^{q}_{s}\;+\;
\delta^{q}_{s}\;\epsilon^{p}_{r}\;}{N\;-\;1}\;+\;\langle pq|rs\rangle
\end{equation}

In this formalism the 1- and 2-order reduced density matrices (1- and 2-RDM) \cite{22} and the 
2-order G-particle-hole correlation matrix \cite{23} may be expressed as   
\begin{equation}
^{1}{\rm
D}^t_{v}\;=\;\langle\Phi|\;a^{t \dagger}a_{v}\;|\Phi\rangle,
\label{d1}
\end{equation}
\begin{equation}
^{2}{\rm D}^{ij}_{kl}\;=\;\frac
1{2!}\;\langle\Phi|\;a^{i \dagger}a^{j \dagger}
a_{l}a_{k}\;|\Phi\rangle\
\label{d2}
\end{equation}
and
\begin{equation}
^2{\rm G}^{im}_{lj}=
\;\langle\Phi|\,^{2}{\hat G}^{im}_{lj}\,|\Phi\rangle\;
=\;\sum_{\Phi^{\prime}\neq\Phi}
\;\langle\Phi|\,a^{i \dagger}a_{m}\;\,
|\Phi^{\prime}\rangle\langle\Phi^{\prime}|
\;a^{j \dagger}a_{l}\,|\Phi\rangle.
\end{equation}
These three matrices, which may be related as follows \cite{23b}
\begin{equation}
2!\;^{2}{\rm D}^{ij}_{ml}\;=\;^{1}{\rm D}^i_{m}\;^{1}{\rm D}^j_{l}\;
-\;^{1}{\rm D}^i_{l}\;{\delta}^j_{m}\;+\;^2{\rm G}^{im}_{lj}
\label{d2g2}
\end{equation}
are at the center of the GHV and HO methodologies.
\vskip 2mm
\noindent
{\bf 2.2. The G-particle-hole hypervirial equation method}
\vskip 2mm

By applying a matrix-contracting mapping involving the G-particle-hole operator  $^{2}{\hat G}$ to the
matrix representation of a particular case of the quantum Liouville equation - the hypervirial of the $N$-electron density operator - one obtains the
GHV equation \cite{9,10}, whose compact form is

\begin{eqnarray}
\left\langle \Phi\,\left|\left\lbrack\hat H,\,
^2{\hat G}^{im}_{lj}\right\rbrack\right|\,\Phi\,\right\rangle\;=\;0
\qquad\qquad(\forall\; i,j,l,m)
\end{eqnarray}
When developing this relation one obtains its explicit form, \cite{9}
\begin{equation}
\begin{split}
&
\sum_{p,q,r,s}\;
^{0}{\rm H}^{rs}_{pq}\;^{(3;2,1)}{\rm C}^{pqj}_{rsl}\;^{1}{\rm D}^{i}_{m}\;
-\;
\sum_{p,q,r,s}\;^{0}{\rm H}^{pq}_{rs}\;^{(3;2,1)}{\rm C}^{rsm}_{pqi}\;^{1}{\rm D}^{l}_{j}\,\\
&+\,2\;
\sum_{p,r,s}\;
^{0}{\rm H}^{rs}_{pm}\;^{(3;2,1)}{\rm C}^{ipj}_{rsl}\;
+\;2\;
\sum_{p,q,r}\;
^{0}{\rm H}^{pq}_{jr}\;^{(3;2,1)}{\rm C}^{lrm}_{pqi}
\\
&+\;2\;
\sum_{p,q,r}\;
^{0}{\rm H}^{ir}_{pq}\;^{(3;2,1)}{\rm C}^{pqj}_{mrl}\;
+\;2\;
\sum_{q,r,s}\;
^{0}{\rm H}^{ql}_{rs}\;^{(3;2,1)}{\rm C}^{rsm}_{jqi}\;=\;0
\end{split}
\label{ghv2}
\end{equation}
where
\begin{equation}
^{(3;2,1)}{\rm C}^{ijm}_{pqt}=\;\sum_{\Phi^{\prime}\neq\Phi}
\;\langle\Phi|\;a^{i\dagger}\;a^{j\dagger}\;a_{q }\;a_{p}\;\,
|\Phi^{\prime}\rangle\langle\Phi^{\prime}|
\;a^{m\dagger}\;a_{t}\;|\Phi\rangle
\end{equation}
are the elements of a 3-order correlation matrix \cite{24}. 

Despite the GHV equation depends not only on 1- and 2-order matrices
but also on 3-order ones, these last matrices can be approximated in terms of the lower-order ones
\cite{8,12,19,24b,29,31b,31c}. The approximation algorithm which is now being used is a recently published modification of
Nakatsuji-Yasuda's one \cite{12,29}.
Proceeding in this way, the solution of the GHV equation may be obtained 
by iteratively solving a set of differential equations to minimize the 2-order
error matrix resulting from the deviation from exact fulfilment of the equation \cite{11}. As a result,
an approximated G-particle-hole matrix corresponding to the eigenstate being considered is obtained \cite{11}.

\vskip 2mm
\noindent
{\bf 2.3. The Hermitian operator method}
\vskip 2mm

In 1973, Bouten, Van Leuven, Mihailovich and Rosina studied the properties
of the particle-hole subspace of a state, and reported the so-called Hermitian Operator method \cite{16,17}, 
which allows one to compute the set of low-lying excited states of an electronic system
from the sole knowledge of the G-particle-hole matrix corresponding to the ground state.
The method is based on a relation connecting the ground state $\Phi$ (reference) with
an excited eigenstate $\Psi$ of the Hamiltonian through an excitation operator ${\hat{\cal S}}$:

\begin{equation}
{\hat H}\,{\hat{\cal S}}\,|\Phi\,\rangle\,=\,E_{\Psi}\,|\Psi\rangle
\label{sch}
\end{equation}
This relation implies the following equivalent equation

\begin{equation}
\langle\,\Phi\,|\,[\,\hat {\cal S,}\,[\,\hat H\,,\,\hat {\cal S}^{\prime}\,]]
|\,\Phi\,\rangle\;=\;(\,E_{\Phi}\,  -\,E_{\Psi}\,)\,\langle\,\Phi\,
|\,\hat {\cal S}\,\hat {\cal S}^{\prime}\,+\,\hat {\cal S}^{\prime}\,\hat {\cal S}\,
|\,\Phi\,\rangle
\label{ghv}
\end{equation}
which has to be solved. To this aim, the authors proposed to approximate the excitation operator as follows, \cite{16}
\begin{equation}
{\hat{\cal S}} = \sum_{t,v}  \{\,c^{(+)}_{t,v}(\,a^{t \dagger} a_{v}
- ^{1}{\rm D}^t_{v} + a^{v \dagger}\;a_{t} - ^{1}{\rm D}^v_{t})
 +\,\,i\,c^{(-)}_{t,v}(\,a^{t \dagger}\,a_{v}\,
-\,^{1}{\rm D}^t_{v} -\,a^{v \dagger}\;a_{t}\,+\,^{1}{\rm D}^v_{t})  \}
\label{ten}
\end{equation}
where the $c$ symbols represent real coefficients and $i$ is the imaginary unit.

By replacing this definition into eq.~(\ref{ghv}), one obtains the following system of decoupled 
equations for the excitation energies $(\,E_{\Phi}\,  -\,E_{\Psi})$ and the expansion
vectors $c^{(\pm)}$

\begin{eqnarray}
\begin{array}{ll}
{\cal H}^{(\pm\pm)}\;c^{(\pm)}&=\,2\;(\,E_{\Psi}\,
  -\,E_{\Phi}\,)\; {\cal G}^{(\pm\pm)}\; c^{(\pm)}
\label{sistema}
\end{array}
\end{eqnarray}
where ${\cal G}^{(\pm\pm)}$ are functionals of the $G$-particle-hole matrix corresponding to the reference eigenstate

\begin{equation}
{\cal G}^{ij (\pm\pm)}_{pq} =\; ^2{\rm G}^{ij}_{pq} \,\pm\,^2{\rm G}^{ij}_{qp}
\,\pm\,^2{\rm G}^{ji}_{pq}\,+\,^2{\rm G}^{ji}_{qp}
\end{equation}
and the matrices ${\cal H}^{(\pm\pm)}$ have the following form

\begin{equation}
\begin{split}
{\cal H}^{ij (\pm\pm)}_{pq}\;&=
\,4\;\sum_{r,s}\;\left\{\, {\tilde {\rm H}}^{jr}_{ps} \,^2{\rm D}_{ir}^{qs}\,
\pm\, {\tilde {\rm H}}^{ir}_{ps} \,^2{\rm D}_{jr}^{qs}\,
\pm\, {\tilde {\rm H}}^{jr}_{qs} \,^2{\rm D}_{ir}^{ps}\,
+\, {\tilde {\rm H}}^{ir}_{qs}  \,^2{\rm D}_{jr}^{ps}\,\right\}\\
&-\,2\;\sum_{r,k,l}\;\left\{\,\delta^q_{i}\, {\tilde {\rm H}^{pr}_{kl}} \,^2{\rm D}^{kl}_{jr}\,
\pm\,\delta^q_{j}\, {\tilde {\rm H}^{pr}_{kl}} \,^2{\rm D}^{kl}_{ir}\,
\pm\,\delta^p_{i}\, {\tilde {\rm H}^{qr}_{kl}} \,^2{\rm D}^{kl}_{jr}\,
+\,\delta^p_{j}\, {\tilde {\rm H}^{qr}_{kl}} \,^2{\rm D}^{kl}_{ir}\,\right\}
\\
&+\,2\;\sum_{k,l}\;\left\{\,{\tilde {\rm H}^{pi}_{kl}} \,^2{\rm D}_{kl}^{jq}\,
\pm\, {\tilde {\rm H}^{pj}_{kl}} \,^2{\rm D}_{kl}^{iq}\,
\pm\, {\tilde {\rm H}^{qi}_{kl}} \,^2{\rm D}_{kl}^{jp}\,
+\,{\tilde {\rm H}^{qj}_{kl}} \,^2{\rm D}_{kl}^{ip}\,\right\}\qquad
\label{mul}
\end{split}
\end{equation}
with
\begin{equation}
{\tilde {\rm H}}^{ir}_{ps}\;=\; ^{0}{\rm H}^{ir}_{ps}\;-\;^{0}{\rm H}^{ri}_{ps}\;\equiv\;
^{0}{\rm H}^{ir}_{ps}\;-\;^{0}{\rm H}^{ir}_{sp}
\label{hami}
\end{equation}
As can be appreciated, the generalized eigenvalue system eq.~(\ref{sistema}) depends only on the 2-RDM, or equivalently
on the G-particle-hole matrix, which happens to be the output of solving the GHV equation. That is why 
we have recently proposed to combine the GHV method with the HO method \cite{18}.
In the following section we outline an algorithm for exploiting point group symmetry, by which the computational efficiency 
of the combined GHV-HO method is highly improved.

\vskip 7mm
\noindent
{\bf 3. Symmetry-adaptation of the GHV-HO method}
\vskip 2mm
 
It is well known that the operations in the symmetry group of a molecule, group ${\cal F}$, maintain the
coefficients of the 2-order electron integral matrix $^0\rm{H}$  unchanged and therefore, this matrix is an invariant (2,2)-tensor
for the group ${\cal F}$ \cite{25}. Analogously, if the $N$-electron state $\Phi$ belongs to a 1-dimensional representation of ${\cal F}$, then 
the 1- and 2-RDM and the G-particle-hole matrix are invariant (1,1)- and (2,2)-tensors for the symmetry group, the formers in the particle-particle metric while the latter in the particle-hole metric \cite{25}. 
Therefore, when the spin-orbitals are 
symmetry-adapted and ordered according to their irreducible representations, these 1- and 2-order matrices
are sparse, and when ${\cal F}$ is Abelian they are also block diagonal. 
The structure of the symmetry forbidden coefficients in all these matrices is easier to analyze when the group ${\cal F}$ 
is an Abelian $D_{2h}$ subgroup, and hence only this kind of groups will be considered hereafter. When the studied electronic system has
non-Abelian symmetry group, an Abelian subgroup will be considered.

The sparsity of all the 1- and 2-order matrices have been recently exploited within 
the framework of the GHV method by carrying out a detailed analysis of the matrix operations involved in eq.~(\ref{ghv2}). This analysis
led to a symmetry-adapted formulation of the GHV algorithm which generates significant computational savings in both floating-point 
operations and memory storage \cite{21}. 
Let us now reconsider the analysis for the case of the HO decoupled equations, eq.~(\ref{sistema}). In this case,
three different types of terms need to be calculated,

\begin{equation}
\sum_{r,s}\;{\tilde {\rm H}}^{jr}_{ps} \;^2{\rm D}_{ir}^{qs}\;\equiv \;^{2}\mathrm{Z}_{pi}^{qj}
\label{z}
\end{equation}
\begin{equation}
\sum_{k,l}\;{\tilde {\rm H}}^{pi}_{kl} \;^2{\rm D}_{jq}^{kl}\;\equiv \;^{2}\mathrm{W}_{jq}^{pi}
\label{w}
\end{equation}
and
\begin{equation}
\sum_{r,k,l}\; \delta^q_i\, {\tilde {\rm H}^{pr}_{kl}} \,^2{\rm D}^{kl}_{jr}\;=\delta^q_i \;^{1}\mathrm{Y}_j^p\;
\;\equiv\;^{2}\mathrm{X}_{ij}^{qp}
\label{x}
\end{equation}
with the auxiliary matrix $^{1}\rm{Y}$ defined as

\begin{equation}
^{1}\mathrm{Y}_j^p \;\equiv \; \sum_{r,k,l}{\tilde {\rm H}^{pr}_{kl}} \,^2{\rm D}^{kl}_{jr}
\label{y}
\end{equation}
A detailed analysis of the mathematical operations involved in the calculation of these terms
reveals that the corresponding auxiliary and final matrices are defined
by covariant equations in particle-particle or particle-hole metric, as appropriate. Those matrices can be expressed in terms of elementary 
tensorial operations as follows:

\begin{equation}
^{2}\mathrm{Z}
= 
\left(\left(
\left({\tilde {\rm H}} \,\otimes\, \;^{2}\mathrm{{\bf D}}\right)
_{\left(1,2,3,4\right)\to\left(3,1,2,4\right)  }^{\left(1,2,3,4\right)\to\left(1,3,4,2\right)}
\right)_{\mathrm{con}}\right)_{\mathrm{con}}
\end{equation}

\begin{equation}
^{2}\mathrm{W}
= 
\left(\left(
\left({\tilde {\rm H}} \,\otimes\, \;^{2}\mathrm{{\bf D}}\right)
_{\left(1,2,3,4\right)\to\left(1,2,3,4\right)  }^{\left(1,2,3,4\right)\to\left(3,4,1,2\right)}
\right)_{\mathrm{con}}\right)_{\mathrm{con}}
\end{equation}

\begin{equation}
^{1}\mathrm{Y}
=
\left(\left(\left(
\left({\tilde {\rm H}} \,\otimes\, \;^{2}\mathrm{{\bf D}}\right)
_{\left(1,2,3,4\right)\to\left(3,4,1,2\right)  }^{\left(1,2,3,4\right)\to\left(1,2,3,4\right)}
\right)_{\mathrm{con}}\right)_{\mathrm{con}}\right)_{\mathrm{con}}
\end{equation}

\begin{equation}
^{2}\mathrm{X} ={\delta}\,\otimes\,\;^{1}\mathrm{{\bf Y}} 
\end{equation}
where
\begin{equation}
\left(\mathrm{{\bf V}}\otimes\mathrm{{\bf W}}\right)  _{m_{1}\ldots m_{v+w}}%
^{i_{1}\ldots i_{v+w}}   =\mathrm{V}_{m_{1}\ldots m_{v}}^{i_{1}\ldots i_{v}%
}\times\mathrm{W}_{m_{v+1}\ldots m_{v+w}}^{i_{v+1}\ldots i_{v+w}}%
\end{equation}

\begin{equation}
\left(\mathrm{{\bf V}}_{\left(1,\ldots,v\right)\to\left(
{\sigma\left(1\right)  },\ldots ,{\sigma\left(
v\right)  }
\right)}^{\left(1,\ldots,v\right)\to\left
({\tau\left(1\right)  },\ldots ,{\tau\left(
v\right)  }
\right)
}\right)  _{m_{1}\ldots m_{v}}^{i_{1}\ldots
i_{v}}   =\mathrm{V}_{m_{\sigma\left(1\right)  }\ldots m_{\sigma\left(
v\right)  }}^{i_{\tau\left(1\right)  }\ldots i_{\tau\left(v\right)  }%
}
\end{equation}

\begin{equation}
\left(\mathrm{{\bf V}}_{\mathrm{con}}\right)  _{m_{1}\ldots m_{v-1}}^{i_{1}\ldots
i_{v-1}}   =\sum_{x}\mathrm{V}_{m_{1}\ldots m_{v-1}x}^{i_{1}\ldots i_{v-1}%
x}
\end{equation}

The covariance of these equations implies that all the intermediate and final matrices
involved in HO method are invariant tensors for the group ${\cal F}$, which retain symmetry properties of the input density and electron integral
matrices. The block structure of these tensors can be applied to efficiently perform the evaluation of the HO
operations for each of the auxiliary operations resulting from eq.~(\ref{sistema}). Thus, for instance, the
auxiliary matrix $^2{\rm Z}$ defined in eq.~(\ref{z}) is a (2,2)-tensor for the group ${\cal F}$ whose 
non-vanishing blocks are associated with irreducible representations $\pi_{i},\pi_{j},\pi_{p},\pi_{q}$
of $\mathcal{F}$ such that $\pi_{i}\otimes\pi_{j}\otimes\pi_{p}\otimes\pi_{q}$=$A$.
Hence, one could avoid the evaluation of the symmetry forbidden elements, and calculate the
remaining elements as follows:

\begin{eqnarray}
^{2}\mathrm{Z}_{pi}^{qj}=\sum_{\substack{\pi_{r},\pi_{s}\\\pi_{j}\otimes
\pi_{r}\otimes\pi_{p}\otimes\pi_{s}=A\\\pi_{i}\otimes\pi_{r}\otimes\pi_{q}\otimes\pi_{s}=A%
}}\sum_{r\in\pi_{r},s\in\pi_{s}}\;{\tilde {\rm H}}_{ps}^{jr}%
\;^{2}\mathrm{D}_{ir}^{qs}\quad(\forall\; p\in\pi_{p},\;q\in
\pi_{q},\;i\in\pi_{i},\;j\in\pi_{j})
\label{symmterm}
\end{eqnarray}
In a similar way, the auxiliary matrix $^2{\rm W}$ defined in eq.~(\ref{w}) can be evaluated as follows:
\begin{equation}
^{2}\mathrm{W}_{jq}^{pi}=\sum_{\substack{\pi_{k},\pi_{l}\\\pi_{p}\otimes
\pi_{i}\otimes\pi_{k}\otimes\pi_{l}=A\\\pi_{k}\otimes\pi_{l}\otimes\pi_{j}\otimes\pi_{q}=A%
}}\sum_{k\in\pi_{k},l\in\pi_{l}}\;{\tilde {\rm H}}^{pi}_{kl}%
\;^{2}\mathrm{D}_{jq}^{kl}\quad(\forall\; p\in\pi_{p},\;q\in
\pi_{q},\;i\in\pi_{i},\;j\in\pi_{j})
\end{equation}
On the other hand, the non-vanishing blocks of elements $^{1}\mathrm{Y}_{p}^{j}$ in eq.~(\ref{y})
 are associated with irreducible representations $\pi_{p},\pi_{j}$ of $\mathcal{F}$
such that $\pi_{p}\otimes\pi_{j}=A$, and for each of these
blocks one calculates
\begin{equation}
^{1}\mathrm{Y}_{p}^{j}=\sum_{\substack{\pi_{r},\pi_{k},\pi_{l}\\\pi_{p}\otimes
\pi_{r}\otimes\pi_{k}\otimes\pi_{l}=A\\\pi_{k}\otimes\pi_{l}\otimes\pi_{j}\otimes\pi_{r}=A%
}}\sum_{r\in\pi_{r},k\in\pi_{k},l\in\pi_{l}}\;{\tilde {\rm H}^{pr}_{kl}} \,^2{\rm D}^{kl}_{jr}\quad(\forall\;p\in\pi_{p},\;j\in
\pi_{j})
\end{equation}

The remaining matrix operations involved in the calculation and solution of the symmetry-blocked HO generalized eigenvalue equations 
can be analyzed and evaluated in a similar way. Therefore, it is possible to exploit the block structure of the ordinary density and electron
integral matrices entering in the HO equations to improve the efficiency of the HO computations and reduce the
memory requirements. In the next Section the computational advantages of a symmetry-adapted
formulation of the GHV-HO (sa-GHV-HO) method, which results from combining the symmetry-adapted formulations of the
GHV (sa-GHV) and HO (sa-HO) algorithms, will be discussed and analyzed.

\vskip 7mm
\noindent
{\bf 4. Results and discussion}
\vskip 2mm

To illustrate the computational advantages of the sa-GHV-HO method, we have carried out a number of calculations on
small to medium sized molecular systems in their ground states at equilibrium experimental geometries \cite{26} in  minimal STO-3G and 
non-minimal 6-31G and 6-31G(d) basis sets. These systems have been chosen in order to explore the computational improvements implemented by the algorithms in different point groups.
The electron integrals for the sa-GHV and sa-HO methods as well as the initial values, at a mean-field level of approximation, of all the matrices 
required for initiating the iterative GHV process have been computed with PSI3 \cite{27}. In order to fairly assess the performance improvement 
due to symmetry, two sets of calculations have been carried out using the same algorithms. Thus, in one set of calculations
we have assumed a C$_{1}$ symmetry group, and in another set the group assumed corresponds to 
the largest Abelian subgroup of the point group describing the full symmetry of
the system determined by PSI3. Consequently, the gains due to symmetry directly reflect the savings inherent in the symmetry-adapted method.

Table 1 reports the statistics pertaining to the computational cost and hardware requirements
of HO calculations. Due to strong dependence on hardware facilities, the tables document
the ratios of the computer time and memory requirements between the calculations performed in
the largest Abelian subgroup of the point group describing the full symmetry of the system determined
by PSI3 and those performed in C$_{1}$ symmetry. As can be appreciated from the documented
data presented in Table 1, the improvement increases not only with the order of the group
but also with the size of the basis set considered. The results show that
computational efficiency ranges from 3.11 to 52.52 in floating-points operations rates and from 1.88 to
7.30 in memory allocation. These computed factors of reduction due to symmetry are indeed close
to the theoretical estimates in most of the cases. Thus, considering that the group $\mathcal{F}$ 
has $f$ irreducible representations, and assuming that
the partitioning of molecular spin-orbitals according to irreducible representation is strictly regular, then
a straightforward calculation shows that $\left(2,2\right)$-tensors have $f$
blocks of size $K^{2}/f\times K^{2}/f$, so they have $K^{4}/f$ non-vanishing
coefficients, and the operations involved in calculation and solution of the generalized eigenvalue equations eq.~(\ref{sistema}) have a time
proportional to $f\times\left(  K^{2}/f\right)  ^{3}=K^{6}/f^{2}$. 
As in the GHV method \cite{21}, these estimates show that the computational costs of the HO method can be
reduced by as much as a factor of $f$ in storage and $f^2$ in floating-point operations.
The asymptotic $f$ and $f^2$ value are only actually achieved when the symmetry blocking
of the orbitals is optimum as can be appreciated from the documented data presented in
Table 1 for the methane molecule. Note that in cases where the dimension of irreducible representation is far from regular,
values of $\sim 0.3\, f^2$ in computer times and
$\sim 0.7\, f$ in memory are achieved. Such is the case of STO-3G acetylene 
which has 4, 0, 1, 1, 0, 4, 1 and 1 orbitals of $a_g$, $b_{1g}$, $b_{2g}$, $b_{3g}$, $a_u$, $b_{1u}$, $b_{2u}$ and $b_{3u}$ symmetries respectively.

\newpage

\vskip 7mm
\noindent
{\bf 5. Concluding remarks}
\vskip 2mm

In this paper, we have outlined a scheme for including the point group symmetry
in GHV-HO calculations. The algorithm provides a means for exploiting sparsity in
the matrices involved in the calculations due to symmetry and
is amenable to an efficient computational implementation. 
The cpu and memory requirements for calculations using this approach
are not limited by the total number of spin-orbitals forming the basis set but rather
by the maximum number of spin-orbitals belonging to the irreducible representations
of the point group describing the full symmetry of the system. Hence, highly symmetric 
large molecules no longer represent a 
formidable computational obstacle. When our implementation of
the sa-GHV-HO method is completed, we plan to apply this technique to studies of challenging examples such as torsional ground- and excited-state potentials in ethylene.
Finally, let us remark that the reported strategy for exploiting symmetry within the GHV-HO
method may also greatly accelerate other RDM-oriented approaches such as the  
contracted Schr\"odinger equation method \cite{7,8,28,29,31,32,33} and the equation-of-motion techniques \cite{18,19,20,34,35,36,37,37b,38}.

\vskip 7mm
\noindent
{\bf Acknowledgements}
\vskip 2mm

This report has been financially 
supported by the Projects UBACYT 20020100100197 and 20020100100502
(Universidad de Buenos Aires, Argentina), PIP N. 11220090100061,
11220090100369 and 11220080100398 (Consejo Nacional
de Investigaciones Cient\'{\i}ficas y T\'ecnicas, Argentina), DI-407-13/I(Universidad Andres Bello, Chile), and PPM12/05, 
GIU12/09 and UFI11/07 (Universidad del Pais Vasco). We thank the 
Universidad del Pais Vasco for allocation of computational resources.


\newpage

\clearpage
\begin{table}[t]
\caption{Comparison of floating-point operations and memory (in brackets)
requirements of the HO computational algorithms:
ratios of the non-symmetry-adapted to the symmetry-adapted formulations.}
\begin{center}
\begin{tabular}{lccccc}
\hline\hline\\[-2em]
System     &Subgroup    &Irr. Rep.&\multicolumn{3}{c}{Basis Set} \\
\cline{4-6}
           &         &   &STO-3G &6-31G  &6-31G(d) \\
\hline
NH$_3$     &C$_s$    &2  &3.11    &3.29    &3.75 \\
           &         &   &[1.88]  &[1.91]  &[1.93]\\
H$_2$O$_2$ &C$_2$    &2  &4.05    &4.27    &3.92 \\
           &         &   &[2.00]  & [2.00] & [2.00]\\
FH         &C$_{2v}$ &4  &4.78    &6.80    &9.36 \\
           &         &   &[2.84]& [3.00]& [3.43]\\
H$_2$O     &C$_{2v}$ &4  &6.05    &8.32    &11.78 \\
           &         &   &[3.09]& [3.20]& [3.54]\\
CH$_4$     &D$_2$    &4  &10.74   &14.34   &15.61 \\
           &         &   &[4.00]& [4.00]& [4.00]\\
C$_2$H$_6$ &C$_{2h}$ &4  &13.72   &20.70   &18.41 \\
           &         &   &[3.76]& [3.82]& [3.87]\\
Li$_2$     &D$_{2h}$ &8  &17.17   &30.86   &47.97 \\
           &         &   &[6.35]& [6.72]& [7.30]\\
C$_2$H$_2$ &D$_{2h}$ &8  &17.55   &24.38   &46.21 \\
           &         &   &[5.68]& [6.00]& [6.87]\\
C$_2$H$_4$ &D$_{2h}$ &8  &21.61   &38.52   &52.52 \\
           &         &   &[6.18]& [6.39]& [7.07]\\
\hline\hline
\end{tabular}
\label{memoryrelat}
\end{center}
\end{table}

\end{document}